\documentclass{article}

\usepackage{microtype}
\usepackage{graphicx}
\usepackage{subcaption}
\usepackage{booktabs}
\usepackage{multirow}
\usepackage{amsmath}
\usepackage{amssymb}
\usepackage{mathtools}
\usepackage{array}
\usepackage{makecell}
\usepackage{enumitem}
\usepackage{xspace}
\usepackage{url}
\usepackage{hyperref}

\usepackage{graphicx}
\usepackage{tikz}
\usepackage{pgfplots}
\pgfplotsset{compat=1.18}

\usepackage[accepted]{icml2026}

\usepackage{amsmath}
\usepackage{amssymb}
\usepackage{mathtools}
\usepackage{amsthm}

\usepackage[capitalize,noabbrev]{cleveref}

\theoremstyle{plain}

\theoremstyle{definition}

\theoremstyle{remark}

\usepackage[textsize=tiny]{todonotes}

\newcommand{\stark}{STaRK\xspace}
\newcommand{\tb}{TraceBound\xspace}
\newcommand{\qwen}{Qwen3\xspace}
\newcommand{\mrr}{MRR\xspace}
\newcommand{\hit}[1]{Hit@#1}
\newcommand{\recall}[1]{Recall@#1}

\icmltitlerunning{When Thinking Before Retrieval Hurts}

\begin{document}

\makeatletter
\renewcommand{\ICML@appearing}{\textit{Accepted at the Failure Modes in Agentic AI (FAGEN) Workshop at ICML 2026}, Seoul, South Korea. Copyright 2026 by the author(s).}
\makeatother

\twocolumn[
  \icmltitle{When Thinking Before Retrieval Hurts:\\
  TraceBound Diagnostics for Adaptive Knowledge-Graph Retrieval}

  \icmlsetsymbol{equal}{*}

  \begin{icmlauthorlist}
    \icmlauthor{Partha Sarathi Purkayastha}{ethz}
  \end{icmlauthorlist}

  \icmlaffiliation{ethz}{Department of Computer Science (D-INFK), ETH Zürich, Zürich, Switzerland}

  \icmlcorrespondingauthor{Partha Sarathi Purkayastha}{ppurkayastha@ethz.ch}

  \icmlkeywords{retrieval agents, knowledge graphs, failure diagnostics, trace evaluation, personalization}

  \vskip 0.3in
]

\printAffiliationsAndNotice{}

\begin{abstract}
Adaptive retrieval promises to make knowledge-graph question answering more robust by letting a controller search, inspect neighborhoods, revise actions, and stop when evidence is sufficient. We study this premise by introducing \textbf{\tb}, a lightweight profile- and trace-conditioned diagnostic protocol for an ARK-style retriever on text-rich knowledge graphs. \tb exposes a compact query profile before retrieval, issues short trace hints after observable failure symptoms, and logs trajectory counters, while keeping graph data, tools, gold labels, and ranking metrics fixed. Across \stark validation and held-out subsets, the added conditioning improves inspectability but consistently reduces retrieval quality under open-weight controllers. Paired trajectory analysis localizes the degradation to repeated calls, zero-result calls, and misallocated exploration budget, while stricter interaction budgets shorten trajectories without repairing the policy. The result diagnoses the common failure mode in that ``thinking before retrieval'' must be evaluated as a control problem over action selection, not as a prompt-format change.
\end{abstract}
\section{Introduction}
Retrieval-augmented generation converts a language-model question answering problem into a joint modeling and retrieval problem where the model must not only reason over evidence, but also decide which evidence should be exposed in the first place. This separation becomes especially consequential for text-rich knowledge graphs. A relevant answer may be recoverable from node descriptions, relation paths, type constraints, or a combination of all three. A natural response is to replace one-shot retrieval with an adaptive retrieval agent where the controller searches broadly, follows promising graph neighborhoods, observes intermediate evidence, and then decides what to retrieve next.

This paper studies a deceptively simple question - When an adaptive graph-retrieval agent is given explicit pre-retrieval ``thinking'' and observation-level trace feedback, does retrieval actually improve? The question matters beyond leaderboard performance. It is also a prerequisite for personalization. A future personalized retriever should use compact user state before retrieval, not only after documents are fetched. However, if profile-like signals are injected into a retrieval controller without a calibrated action policy, they may change the trajectory without improving the ranked answer list.

We introduce \textbf{\tb}, a diagnostic protocol for adaptive knowledge-graph retrieval. \tb has two roles. First, it is an intervention in which the controller receives a compact query profile before acting and receives trace-conditioned hints when the trajectory exhibits symptoms such as repeated calls, zero-result calls, or budget exhaustion. Second, it is an instrumentation layer where every trajectory is converted into process counters that can be paired with final ranking metrics. The protocol is deliberately lightweight. It does not alter graph data, training labels, gold answers, ranking metrics, or the underlying search tools. This makes the comparison strict, so if profile and trace conditioning are useful, they should improve the agent under the same retrieval interface.

The empirical result is strong but not in the direction one might expect. On \stark, \tb makes the retrieval process more inspectable but reduces retrieval quality relative to the unconditioned ARK-style controller in our open-weight setup. The degradation appears in a Qwen3-14B validation ablation, in a Qwen3-30B-A3B held-out subset, and in shared-ID paired comparisons across PRIME, MAG, and Amazon. The trace analysis explains why the result is not merely noise as repeated calls and zero-result calls are associated with substantially larger negative rank deltas, and a stricter budget reduces trajectory length without recovering accuracy.

We organize the study around three research questions. \textbf{RQ1}: does profile/trace conditioning improve standard retrieval metrics when the underlying graph tools and evaluator are fixed? \textbf{RQ2}: if ranking quality changes, can trajectory counters explain the direction of the change? \textbf{RQ3}: does imposing a stricter interaction budget repair the failure, or does the system require a different action policy? These questions make the paper a controlled analysis of adaptive retrieval behavior rather than a comparison of unrelated retriever architectures.

Our contributions are threefold. First, we provide a reproducible profile/trace intervention for adaptive graph retrieval that isolates controller behavior from dataset and metric changes. Second, we report paired ranking and trace diagnostics showing when the intervention hurts. Third, we identify the operational mechanism of failure that profile and trace state are useful for diagnosis, but harmful as unconditional prompt context unless paired with a calibrated retrieval-action policy.

\section{Related Work}
\paragraph{Retrieval-augmented generation and adaptive retrieval.}
Retrieval-augmented generation combines parametric generation with non-parametric evidence access \citep{lewis2020rag}. Dense passage retrieval established a strong neural retrieval paradigm for open-domain question answering \citep{karpukhin2020dpr}, while later systems interleave retrieval with reasoning, reflection, or critique \citep{trivedi2023interleaving,asai2024selfrag}. These methods motivate interactive evidence acquisition, but many assume a document collection rather than a text-rich graph whose structure constrains valid answers.

\paragraph{Language-model agents and tool use.}
Tool-using agents such as ReAct combine reasoning traces with environment actions \citep{yao2023react}. Reflexion, Self-Refine, and Toolformer explore variants of feedback, iterative improvement, and learned tool use \citep{shinn2023reflexion,madaan2023selfrefine,schick2023toolformer}. In principle, such methods suggest that a retrieval controller should improve after seeing observations. Our results identify a boundary condition where observation feedback can make the process more measurable while still degrading final retrieval when the next-action policy is not calibrated.

\paragraph{Graph retrieval over textual and relational knowledge bases.}
Knowledge-graph reasoning methods such as Think-on-Graph expose graph traversal to language models \citep{sun2023thinkongraph}. \stark is the most relevant benchmark for our setting because it evaluates retrieval over semi-structured textual and relational knowledge bases across product search, academic paper search, and precision medicine \citep{wu2024stark}. Autofocus or AF-Retriever studies multi-hop retrieval over semi-structured knowledge bases by combining entity search, LLM-generated Cypher queries, node-set joins, vector search, and reranking \citep{boer2025afretriever}. GraphRAFT explores retrieval-augmented fine-tuning for graph databases \citep{clemedtson2025graphraft}. ARK is the closest architectural reference and it gives a language model two graph tools, global lexical search with one-hop neighborhood exploration, so that the model can balance breadth-oriented discovery with depth-oriented traversal \citep{polonuer2026ark}. We build on the same broad premise but focus on trace-conditioned failure analysis rather than claiming a new state-of-the-art retriever.

\paragraph{Personalization before retrieval.}
Personalized retrieval work argues that user-specific semantics should influence retrieval before documents are fetched. Personalize Before Retrieve (PBR) performs personalized query expansion using style-aware pseudo feedback and corpus-structure alignment \citep{zhang2025pbr}. PersonaBench evaluates whether systems can answer questions requiring access to synthetic personal information \citep{tan2025personabench}. Our experiments do not introduce real user profiles; the profile in \tb is query-derived. The relevance to personalization is methodological as a user profile is another pre-retrieval signal, and our results show that adding such a signal globally to an agent prompt is not a substitute for learning when it should change the retrieval action.

\paragraph{Failure diagnostics for agents.}
Final answer metrics can hide process failures as an agent may make redundant calls, chase empty branches, or recover by chance after an inefficient trajectory. Agent evaluation work therefore increasingly emphasizes trajectories, reproducible failure triggers, and process-level measurements \citep{liu2023agentbench,xie2024osworld}. \tb follows this instrumentation-first view. We pair standard retrieval metrics with trajectory counters so that failure is attributed to observable behavior rather than only to aggregate score changes.

\section{TraceBound: Methodology and Implementation Details}
\subsection{Problem setting and interface}
Each example consists of a natural-language query $q$, a text-rich knowledge graph $G=(V,E)$, and a gold answer set $Y \subseteq V$. A retrieval agent returns an ordered list $\hat{Y}=(v_1,\ldots,v_k)$, which is evaluated with \hit{1}, \hit{5}, \recall{20}, and \mrr. The agent can call two graph tools: \emph{global search}, which retrieves candidate nodes from textual descriptors, and \emph{neighborhood exploration}, which expands a selected node through one-hop relations. The controller operates in an action-observation loop until it emits final answers or reaches a maximum interaction budget.

\subsection{The TraceBound intervention}
\tb adds profile conditioning and trace conditioning to this controller. The input remains the original query and graph; the output remains a ranked node list plus an execution trajectory. The intervention changes the controller context, not the task definition.

\paragraph{Query profile.}
Before the first tool call, \tb asks the controller to form a compact profile containing predicted answer type, important query facets, relation hints, and retrieval risks. The profile is not a user memory; it is a query-derived stand-in for the kind of compact state that a personalized retriever would eventually use. Its purpose is to test whether an explicit pre-retrieval state helps the controller decide how to search.

\paragraph{Trace hints.}
After each observation, \tb updates a trace state with counters for global calls, neighborhood calls, zero-result calls, repeated calls, step count, and remaining budget. The controller receives a short diagnostic hint only when the trace indicates a potential failure pattern, such as repeated identical calls or empty neighborhoods. The hint is intentionally constrained; it can recommend re-anchoring, relaxing a constraint, tightening type focus, or stopping, but it does not reveal labels or gold answers.

\paragraph{Instrumentation.}
Every trajectory is logged as a sequence of actions, observations, and counters. This lets us compute paired process metrics on the same examples used for ranking evaluation. The key design choice is separation - ranking metrics are computed only from final answer lists, while trace metrics are computed only from trajectories.

\begin{figure*}[t]
\centering
\includegraphics[width=\textwidth]{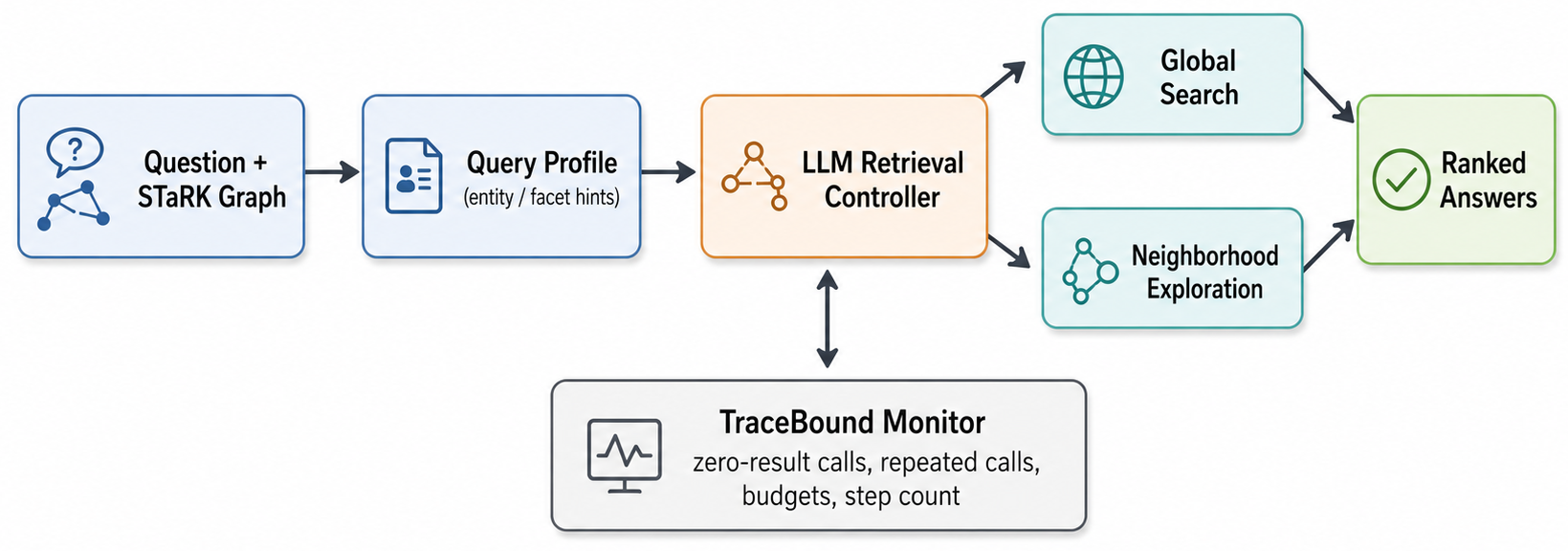}
\caption{\textbf{TraceBound workflow.} The input query and graph are passed to an ARK-style controller. \tb adds a query-profile block before retrieval and a monitor that summarizes zero-result calls, repeated calls, budget use, and step count during retrieval. The graph tools and final ranking evaluator are unchanged. The ranked answers produce \hit{1}, \hit{5}, \recall{20}, and \mrr, while the monitor produces trace diagnostics used only for analysis.}
\label{fig:pipeline}
\end{figure*}

\subsection{Evaluated controller variants}
Table~\ref{tab:modes} defines the four variants. The unconditioned agent is the baseline: it receives the original query, tool descriptions, and prior observations, but no additional profile or trace hint. Profile-only and trace-only isolate the two ingredients. \tb combines both.

\begin{table}[t]
\centering
\caption{\textbf{Controller variants.} ``Profile'' means the controller receives a compact pre-retrieval query profile. ``Trace hint'' means the controller receives short observation-level feedback derived from trajectory counters. ``Logged diagnostics'' means the trajectory is converted into process metrics; these metrics are not used by the final ranking evaluator.}
\label{tab:modes}
\resizebox{\columnwidth}{!}{%
\begin{tabular}{lccc}
\toprule
Variant & Profile & Trace hint & Logged diagnostics \\
\midrule
Unconditioned agent & no & no & final-answer metrics \\
Profile-only agent & yes & no & profile metadata \\
Trace-only agent & no & yes & process counters \\
\tb agent & yes & yes & profile + process counters \\
\bottomrule
\end{tabular}}
\end{table}

\subsection{Algorithmic summary}
Given a query $q$, \tb first constructs a query profile $p(q)$. At step $t$, the controller observes the query, profile, previous observations, and current trace summary $s_t$. It chooses either global search, neighborhood exploration, or final-answer emission. After a tool call, the monitor updates $s_{t+1}$ by checking whether the result was empty, whether the call repeated an earlier call, how much budget remains, and whether enough candidates have been accumulated. The method is therefore not a new retriever index; it is a controller-side protocol for profile-conditioned action selection and trace-conditioned diagnostics.

\subsection{Design invariants}
A diagnostic intervention is only useful if it does not move the evaluation target. We therefore preserve five invariants. Table~\ref{tab:invariants} summarizes what is fixed and what is changed. The graph corpus, gold labels, retrieval metrics, and tool semantics remain fixed. The only manipulated object is the controller context and the logging attached to each action. This lets us attribute changes in ranking to the behavior of the adaptive policy rather than to a different retriever, different corpus, or different evaluator.

\subsection{What the trace can and cannot do}
The trace hint is not an oracle. It does not know whether a retrieved node is correct, and it does not inspect the gold set. It only summarizes observable symptoms. For example, if the controller repeats a global search with nearly identical arguments, the hint can recommend changing the anchor or stopping repeated exploration. If a neighborhood call returns no candidates, the hint can recommend relaxing the relation/type constraint or returning to global search. This keeps \tb close to deployment settings where a retrieval system can observe its own trajectory but cannot observe the correct answer during inference.

\subsection{State representation and action view}
The protocol can be written as a small partially observed control problem. At step $t$, the controller has a hidden retrieval belief induced by the prompt and an explicit trace state
\[
  s_t = (g_t, n_t, z_t, r_t, b_t, c_t),
\]
where $g_t$ counts global searches, $n_t$ counts neighborhood explorations, $z_t$ counts zero-result calls, $r_t$ counts repeated calls, $b_t$ is remaining budget, and $c_t$ summarizes the current candidate set. The controller chooses an action $a_t \in \{\textsc{Search},\textsc{Explore},\textsc{Answer}\}$. \tb changes the textual representation of $s_t$ exposed to the controller, while the evaluator observes only the final $\textsc{Answer}$ action. This view is useful because it makes clear why static prompts are limited: the important object is the conditional distribution over actions, not the mere presence of a trace string.

\section{Experimental Setup}
\paragraph{Benchmark and metrics.}
We evaluate on \stark PRIME, MAG, and Amazon. These graphs cover biomedical entities, academic papers, and products, respectively. We report the standard retrieval metrics \hit{1}, \hit{5}, \recall{20}, and \mrr. \hit{1} and \hit{5} measure whether at least one correct answer appears in the first one or five ranks. \recall{20} measures the fraction of gold answers recovered in the first twenty ranks. \mrr measures the reciprocal rank of the first correct answer and is sensitive to whether the correct node moves near the top of the list.

\paragraph{Models and serving.}
All experiments use open-weight \qwen controllers \cite{yang2025qwen3} served through vLLM. Validation ablations use Qwen3-14B. Held-out subset and constrained-budget experiments use Qwen3-30B-A3B-Instruct-2507 with tensor parallel serving. This local serving choice is reported for reproducibility and cost transparency; the research question is about adaptive retrieval behavior, not about a particular deployment constraint.

\paragraph{Evaluation protocol.}
The validation ablation uses 100 examples per graph and compares four controller variants under two maximum interaction budgets: 12 and 16. The held-out subset uses the stronger controller and compares the unconditioned agent against \tb on PRIME, MAG, and Amazon. Because long-running agent jobs can complete slightly different numbers of examples per condition, we report aggregate results and shared-ID paired results. Paired analysis is the most conservative comparison as it keeps only query IDs present in both conditions and computes \tb-minus-unconditioned deltas on the same examples.

\paragraph{Constrained-budget probe.}
To test whether degradation is caused mainly by excessive interaction length, we run a constrained-budget version of \tb with a lower step budget and tighter caps on global and neighborhood calls. This probe is not a separate model; it is a controlled change to the same profile/trace protocol. Its purpose is to distinguish ``the agent thinks too long'' from ``the agent chooses the wrong actions.''

\paragraph{Implementation.}
The implementation overlays \tb on an ARK-style code path where graph tools, answer formatting, and metric scripts are preserved, while prompts, trace logging, and run wrappers are extended. We avoid training or test-set tuning since the contribution is evaluated as a controller protocol over fixed tools and fixed metrics.

\paragraph{Naming convention.}
We avoid shorthand labels in the analysis. ``Unconditioned'' denotes the original controller without profile or trace hints. ``Profile-only'' and ``Trace-only'' are single-component ablations. ``\tb'' denotes the combined profile-plus-trace controller and its instrumentation. A label such as ``Unconditioned, 12 steps'' therefore means that the same controller is run with a maximum of 12 action-observation iterations; it does not refer to a separate model checkpoint.

\paragraph{Evidence structure.}
The reported tables are produced from JSON trajectory logs and metric summaries. Each per-query log contains the final ranked list, tool calls, observations, and derived counters. The package includes the table summaries and paired-diagnostic outputs used to construct the paper. The raw traces make it possible to audit whether a score difference arises from a ranking change, a missing completion, or a trajectory-level failure such as a repeated call.

\section{Results}
The central empirical finding is consistent across model scale, graph domain, and comparison method - profile and trace conditioning make trajectories easier to analyze, but they do not improve final ranking in this implementation.

\subsection{Validation ablation: profile and trace conditioning hurt MRR}
Table~\ref{tab:validation} reports the Qwen3-14B validation ablation averaged across PRIME, MAG, and Amazon. The ``Steps'' column is the maximum number of action-observation iterations allowed before final-answer emission. The metric columns are averages across the three graphs. The unconditioned agent with a 12-step budget obtains the best average \mrr (.4087), while the complete \tb agent obtains .3141 at 12 steps and .3103 at 16 steps. The profile-only and trace-only variants also underperform the unconditioned agent, showing that the degradation is not caused by only one component.

\begin{table}[t]
\centering
\caption{\textbf{Validation ablation averaged over PRIME, MAG, and Amazon.} The unconditioned agent is strongest, while the combined \tb agent loses both rank precision and recall.}
\label{tab:validation}
\resizebox{\columnwidth}{!}{%
\begin{tabular}{lrrrrr}
\toprule
Controller variant & Steps & \hit{1} & \hit{5} & \recall{20} & \mrr \\
\midrule
Unconditioned agent & 12 & \textbf{.3333} & \textbf{.5100} & \textbf{.4252} & \textbf{.4087} \\
 & 16 & .3233 & .5067 & .4192 & .4058 \\
\midrule
Profile-only agent & 12 & .2833 & .4533 & .3845 & .3511 \\
 & 16 & .2533 & .4700 & .3848 & .3473 \\
\midrule
Trace-only agent & 12 & .2333 & .3767 & .2840 & .2944 \\
 & 16 & .2733 & .4167 & .3195 & .3307 \\
\midrule
\tb agent & 12 & .2579 & .3951 & .3036 & .3141 \\
 & 16 & .2567 & .3900 & .2998 & .3103 \\
\bottomrule
\end{tabular}}
\end{table}

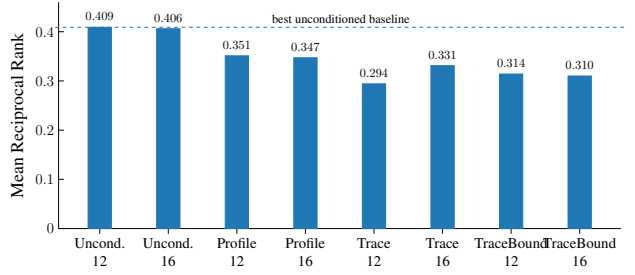
\begin{figure}[t]
\centering
\resizebox{\columnwidth}{!}{\providecolor{mplblue}{HTML}{1F77B4}

\begin{tikzpicture}
\begin{axis}[
    width=6.9in,
    height=3.15in,
    ybar,
    bar width=0.66cm,
    ymin=0,
    ymax=0.46,
    ylabel={Mean Reciprocal Rank},
    xtick={1,2,3,4,5,6,7,8},
    xticklabels={
        {Uncond.\\12},
        {Uncond.\\16},
        {Profile\\12},
        {Profile\\16},
        {Trace\\12},
        {Trace\\16},
        {TraceBound\\12},
        {TraceBound\\16}
    },
    xticklabel style={align=center,font=\large},
    xmin=0.4,
    xmax=8.6,
    ytick={0.0,0.1,0.2,0.3,0.4},
    tick label style={font=\large},
    label style={font=\Large},
    title style={font=\LARGE},
    axis x line*=bottom,
    axis y line*=left,
    axis line style={black,semithick},
    tick style={black},
    nodes near coords={
        \pgfmathprintnumber[fixed,precision=3,zerofill]{\pgfplotspointmeta}
    },
    point meta=rawy,
    every node near coord/.append style={
        font=\normalsize,
        black,
        anchor=south,
        yshift=2pt
    },
    clip=false,
]
\addplot+[fill=mplblue,draw=mplblue] coordinates {
    (1,0.409)
    (2,0.406)
    (3,0.351)
    (4,0.347)
    (5,0.294)
    (6,0.331)
    (7,0.314)
    (8,0.310)
};

\draw[mplblue,dashed,thick]
    (axis cs:0.35,0.409) -- (axis cs:8.65,0.409);

\node[anchor=west,font=\normalsize]
    at (axis cs:3.45,0.426)
    {best unconditioned baseline};

\end{axis}
\end{tikzpicture}}
\caption{\textbf{Validation MRR by controller variant and step budget.} The dashed horizontal reference line marks the best unconditioned baseline. All profile/trace variants fall below that line, so the intervention improves neither the 12-step nor the 16-step validation setting.}
\label{fig:validation_mrr}
\end{figure}

\subsection{Held-out subset: the direction persists with a stronger controller}
Table~\ref{tab:heldout} reports the Qwen3-30B-A3B held-out subset. The ``$n$'' column is the number of completed examples in that graph and condition. The pattern is again negative for \tb. On MAG, the most severe case, \tb reduces \mrr from .7711 to .4974. PRIME and Amazon have smaller but still negative changes. This matters because MAG is the graph where relation-rich academic queries can strongly reward precise anchoring; misdirected adaptive exploration can therefore move the controller toward plausible but wrong neighborhoods.

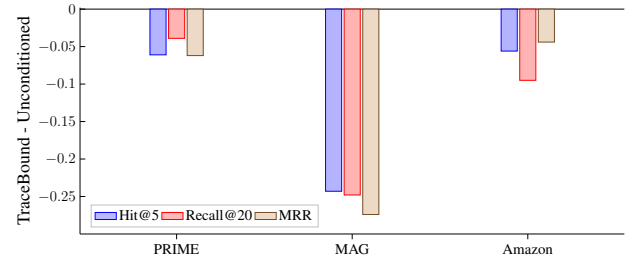
\begin{figure}[t]
\centering
\resizebox{\columnwidth}{!}{\providecolor{mplblue}{HTML}{1F77B4}
\providecolor{mplorange}{HTML}{FF7F0E}
\providecolor{mplgreen}{HTML}{2CA02C}

\begin{tikzpicture}
\begin{axis}[
    width=6.7in,
    height=3.15in,
    ybar,
    area legend,
    bar width=13pt,
    ymin=-0.3,
    ymax=0.002,
    xmin=0.45,
    xmax=3.55,
    ylabel={TraceBound - Unconditioned},
    xtick={1,2,3},
    xticklabels={PRIME,MAG,Amazon},
    ytick={0,-0.05,-0.10,-0.15,-0.20,-0.25},
    yticklabel style={/pgf/number format/fixed,/pgf/number format/precision=2},
    xticklabel style={font=\large},
    tick label style={font=\large},
    label style={font=\Large},
    title style={font=\LARGE},
    axis x line*=bottom,
    axis y line*=left,
    axis line style={black,semithick},
    tick style={black},
    extra y ticks={0},
    extra y tick labels={},
    extra y tick style={
        grid=major,
        major grid style={black,semithick},
        tick style={draw=none}
    },
    legend style={
        at={(0.02,0.03)},
        anchor=south west,
        draw=gray!50,
        fill=white,
        fill opacity=0.92,
        text opacity=1,
        legend columns=3,
        font=\large
    },
    legend cell align={left},
]
\addplot+[] coordinates {
    (1,-0.061)
    (2,-0.243)
    (3,-0.056)
};
\addplot+[] coordinates {
    (1,-0.039)
    (2,-0.248)
    (3,-0.095)
};
\addplot+[] coordinates {
    (1,-0.062)
    (2,-0.274)
    (3,-0.044)
};
\legend{Hit@5, Recall@20, MRR}
\end{axis}
\end{tikzpicture}}
\caption{\textbf{Held-out metric deltas.} Each grouped bar is \tb minus unconditioned on the held-out subset. Bars below zero mean the profile/trace intervention is worse. The three bars per graph correspond to \hit{5}, \recall{20}, and \mrr; all are negative across PRIME, MAG, and Amazon.}
\label{fig:heldout_delta}
\end{figure}

\begin{table}[t]
\centering
\caption{\textbf{Held-out subset with Qwen3-30B-A3B.} Each row is one graph-condition cell. ``$n$'' gives completed examples. The \tb rows underperform the corresponding unconditioned rows on every graph.}
\label{tab:heldout}
\resizebox{\columnwidth}{!}{%
\begin{tabular}{llrrrrr}
\toprule
Graph & Controller & $n$ & \hit{1} & \hit{5} & \recall{20} & \mrr \\
\midrule
PRIME & Unconditioned & 290 & .2966 & .4207 & .3681 & .3499 \\
 & \tb & 300 & .2400 & .3600 & .3297 & .2880 \\
\midrule
MAG & Unconditioned & 299 & .7224 & .8328 & .7483 & .7711 \\
 & \tb & 300 & .4233 & .5900 & .5006 & .4974 \\
\midrule
Amazon & Unconditioned & 296 & .5439 & .7128 & .4557 & .6152 \\
 & \tb & 300 & .5033 & .6567 & .3607 & .5714 \\
\bottomrule
\end{tabular}}
\end{table}

\subsection{Paired comparison: the result is not an artifact of completed-count differences}
Table~\ref{tab:paired} keeps only shared query IDs available in both unconditioned and \tb outputs. ``Paired $n$'' is the number of shared examples. Each delta is \tb minus unconditioned on the same queries. The paired comparison confirms the aggregate result: all reported deltas are negative. MAG again shows the largest effect, with a -.2711 paired \mrr delta.

\begin{table}[t]
\centering
\caption{\textbf{Shared-ID paired diagnostics.} Rows compare \tb and the unconditioned agent on identical query IDs. $\Delta\hit{5}$, $\Delta\recall{20}$, and $\Delta\mrr$ are \tb-minus-unconditioned. Negative values indicate that \tb ranks correct answers lower or retrieves fewer gold answers.}
\label{tab:paired}
\resizebox{0.85\columnwidth}{!}{%
\begin{tabular}{lrrrr}
\toprule
Graph & Paired $n$ & $\Delta\hit{5}$ & $\Delta\recall{20}$ & $\Delta\mrr$ \\
\midrule
PRIME & 291 & -.0653 & -.0421 & -.0653 \\
MAG & 300 & -.2400 & -.2452 & -.2711 \\
Amazon & 297 & -.0539 & -.0911 & -.0426 \\
\bottomrule
\end{tabular}}
\end{table}

\subsection{Effect-size summary}
Across the three held-out graphs, the average paired \mrr delta is negative. The magnitude is not uniform. MAG accounts for the largest drop, PRIME is moderate, and Amazon is smaller but still negative. This heterogeneity is scientifically useful: it shows that adaptive failure depends on graph structure and query type. A controller that can recover on product queries may still fail badly on academic graph queries where relation neighborhoods are dense and plausible distractors are common.

\section{Discussion}

\subsection{Trace metrics identify the mechanism}
Table~\ref{tab:trace} reports process counters for the \tb trajectories in the held-out paired set. The columns should be read as follows: ``Global'' is the average number of broad graph-search calls, ``Neigh.'' is the average number of one-hop neighborhood exploration calls, ``Zero'' is the average number of tool calls returning no useful candidate, ``Repeat'' is the average number of repeated or near-duplicate calls, ``Events'' is the average number of logged trace events, and ``Steps'' is the average controller step count. These counters reveal that \tb often spends substantial budget on exploration that does not translate into better ranking.

\begin{table}[t]
\centering
\caption{\textbf{Trace counters for \tb on held-out paired examples.} All entries are per-query averages. Higher values are not inherently better, they only indicate more interaction budget consumed before final ranking.}
\label{tab:trace}
\resizebox{\columnwidth}{!}{%
\begin{tabular}{lrrrrrr}
\toprule
Graph & Global & Neigh. & Zero & Repeat & Events & Steps \\
\midrule
PRIME & 4.1478 & 5.7010 & 4.6151 & 1.6564 & 11.8179 & 8.2577 \\
MAG & 2.4133 & 5.0200 & 2.2767 & 0.6633 & 9.8333 & 5.8667 \\
Amazon & 2.4040 & 5.4815 & 3.9293 & 0.6364 & 10.3131 & 6.1111 \\
\bottomrule
\end{tabular}}
\end{table}

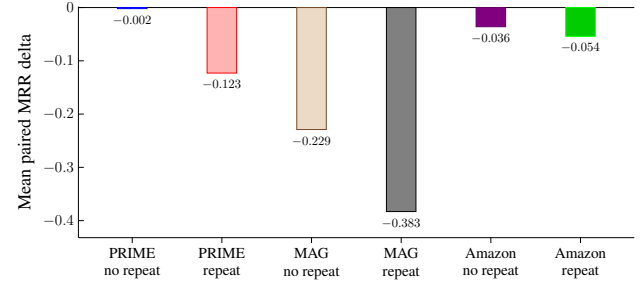
\begin{figure}[t]
\centering
\resizebox{\columnwidth}{!}{\begin{tikzpicture}
\begin{axis}[
    width=6.7in,
    height=3.15in,
    ybar,
    bar width=0.82cm,
    every axis plot/.append style={bar shift=0pt},
    ymin=-0.432,
    ymax=0.002,
    ylabel={Mean paired MRR delta},
    xtick={1,2,3,4,5,6},
    xticklabels={
        {PRIME\\no repeat},
        {PRIME\\repeat},
        {MAG\\no repeat},
        {MAG\\repeat},
        {Amazon\\no repeat},
        {Amazon\\repeat}
    },
    xticklabel style={align=center,font=\large},
    xmin=0.4,
    xmax=6.6,
    ytick={0,-0.1,-0.2,-0.3,-0.4},
    tick label style={font=\large},
    label style={font=\Large},
    title style={font=\LARGE},
    axis x line*=bottom,
    axis y line*=left,
    axis line style={black,semithick},
    tick style={black},
    extra y ticks={0},
    extra y tick labels={},
    extra y tick style={
        grid=major,
        major grid style={black,semithick},
        tick style={draw=none}
    },
    nodes near coords={
        \pgfmathprintnumber[fixed,precision=3,zerofill]{\pgfplotspointmeta}
    },
    point meta=rawy,
    every node near coord/.append style={
        font=\normalsize,
        black,
        anchor=north,
        yshift=-2pt
    },
    clip=false,
]

\addplot+[] coordinates {(1,-0.002)};
\addplot+[] coordinates {(2,-0.123)};
\addplot+[] coordinates {(3,-0.229)};
\addplot+[] coordinates {(4,-0.383)};
\addplot+[] coordinates {(5,-0.036)};
\addplot+[] coordinates {(6,-0.054)};

\end{axis}
\end{tikzpicture}}
\caption{\textbf{Conditional rank loss by repeated-call bucket.} Each bar is the mean paired \mrr delta, \tb minus unconditioned. Here, ``repeat'' means at least one repeated or near-duplicate call occurred. Repetition predicts a larger loss on PRIME and MAG, and does not help on Amazon either.}
\label{fig:repeat_delta}
\end{figure}

The repeated-call counter is especially informative. Figure~\ref{fig:repeat_delta} conditions paired \mrr deltas on whether a \tb trajectory contains any repeated call. Each bar is the mean \tb-minus-unconditioned \mrr delta for a graph and bucket. PRIME is almost neutral when no repeated call occurs (-.0023), but becomes substantially worse when repetition appears (-.1228). MAG is negative in both buckets, yet repetition makes the degradation larger. The result suggests that repetition is not only an efficiency problem; it is a symptom of a controller that has lost the evidence trail.

\subsection{A tighter budget is not sufficient}
The constrained-budget probe asks whether \tb fails simply because it explores too long. Table~\ref{tab:budget} lowers the interaction budget and caps the number of global and neighborhood calls. The ``Steps'' column confirms that the probe shortens trajectories, but the rank metrics remain weak. This separates two hypotheses. If the main failure were only excessive length, tighter budgets should recover most of the unconditioned performance. Instead, the controller remains inaccurate, indicating that the problem is action selection for when to anchor, when to traverse, when to relax, and when to stop.

\subsection{Why the result matters}
The important finding is not simply that a prompt-level intervention failed. It is that the failure is structured and measurable. TraceBound changes the retrieval trajectory in ways that can be audited by showing how repeated calls indicate stagnation, zero-result calls indicate failed anchoring or traversal, and high exploration counts indicate budget spent without rank improvement. These are exactly the kinds of failures that final retrieval metrics can hide. This matters for adaptive graph retrieval because errors compound over the trajectory. A wrong early anchor can send the controller into a plausible but irrelevant neighborhood. A trace hint can then make the system more active around that bad anchor rather than causing a clean reset. In this setting, more explicit reasoning does not automatically produce better retrieval, instead it only creates more state for the controller to manage. The practical lesson is that profile and trace information should not be treated as globally useful context. They should be inputs to a calibrated action policy that decides when to search broadly, traverse locally, relax constraints, rerank, stop, or restart. TraceBound is therefore useful not because it improves the controller directly, but because it exposes the failure signals that a better controller must learn to act on.

\subsection{Failure taxonomy and corrective policy targets}
The trace counters support a taxonomy of operational failures. Table~\ref{tab:taxonomy} lists the failure class, the observable trigger, its likely ranking effect, and the policy correction suggested by the result. The final column is intentionally framed as a policy target rather than a prompt instruction. The experiments show that simply telling the controller about a failure symptom is insufficient; the system must learn when the symptom should change the next retrieval action.

\subsection{Design implications}
The study suggests three concrete design principles. First, a profile should be treated as an uncertain signal, not as privileged context. A controller should be able to ignore it when graph constraints are sufficient. Second, trace feedback should be gated by failure confidence. Repetition and empty calls are useful symptoms, but reacting too early can interrupt a correct path. Third, stopping should be a learned action. A system that only searches or explores will keep consuming budget even after the marginal value of another call is low.

\subsection{Implications for personalization}
The query profile in \tb is not a real user model, but it plays the same architectural role as a user profile would play in a personalized retriever - compact context injected before evidence is fetched. The negative result therefore gives a useful warning. Personalization should not be globally concatenated to every retrieval decision. Instead, the controller should learn a conditional policy which uses the profile when it helps disambiguate intent or rank candidates, ignores it when graph constraints dominate, and backtracks when profile-guided expansion enters an empty branch. In other words, personalization is not only a representation problem, it is a control problem over retrieval actions.

\section{Limitations}
Our evidence is specific to an ARK-style two-tool interface, \stark graphs, and open-weight \qwen controllers served locally. We evaluate query-derived profiles rather than real user memories, so the experiments should not be read as a complete personalized-retrieval benchmark. The held-out evaluation is a subset rather than the full test split, and the constrained-budget probe is diagnostic rather than a tuned production system. These constraints bound our claim that \tb shows that a plausible profile/trace-conditioned controller can degrade adaptive graph retrieval and that trace diagnostics explain the degradation. It, however, does not prove that all adaptive or personalized retrieval interventions will fail.

\section{Future work}
The next step is to replace prompt-level intervention with policy learning. The traces collected here define the state variables: candidate coverage, repeated calls, zero-result calls, relation coverage, budget use, and rank movement. A learned controller can use these variables to choose among global search, neighborhood exploration, relation relaxation, type tightening, reranking, and stopping. Personalization should enter this policy selectively. For example, a user profile may be useful for query anchoring or final reranking but harmful during relation expansion if it distracts from graph constraints. A validation-trained router that decides when to activate profile conditioning is a natural bridge from this diagnostic study to robust personalized retrieval.

\paragraph{A concrete policy-learning formulation.}
A direct extension is to train a small controller or router over the trace state rather than rely on unconditional textual hints. The training instance would consist of the query profile, current candidate summary, trace counters, and the next action that improves a validation objective. The action space can remain small: broad search, neighborhood exploration, relation relaxation, type tightening, rerank, answer, or restart. The reward can combine rank movement, recall gain, and budget cost. This formulation also makes personalization precise. A user state is not appended everywhere; it becomes one feature available to the policy, and the policy is rewarded only when using that feature improves retrieval.

\section{Conclusion}
\tb turns the common intuition that ``more explicit thinking before retrieval should help an adaptive KG retriever" into a measurable experiment. Under our controlled implementation, it does not. The unconditioned agent remains stronger on validation, held-out subsets, and paired shared-ID comparisons. The value of the study is that the reason becomes visible in the fact that repeated calls, zero-result calls, and misallocated exploration budget predict rank loss. This shifts the design target for personalized reasoned retrieval as the central problem is no longer how to add more profile text to the prompt, but how to learn a calibrated retrieval policy that knows when profile and trace information should influence the next action.

\section*{Impact Statement}
This work studies reliability failures in retrieval agents and is intended to improve transparent evaluation before such agents are deployed in personalized or private-data settings. It does not introduce new user data or new automated decision pipelines. The main practical risk is overgeneralization as our experiments only show that one plausible profile- and trace-conditioned controller can degrade retrieval quality under a specific open-weight configuration, not that adaptive retrieval or personalization is intrinsically harmful. Trace reporting like these can help researchers identify brittle retrieval policies before they are embedded in user-facing systems.

\bibliography{example_paper}
\bibliographystyle{icml2026}

\appendix
\onecolumn
\section*{Appendix}

\begin{table*}[h]
\centering
\caption{\textbf{Failure taxonomy induced by trace diagnostics.} ``Trigger'' is computed from the trajectory. ``Effect'' describes how the failure can damage ranking. ``Policy target'' states what a learned controller should decide, rather than what a static prompt should always instruct.}
\label{tab:taxonomy}
\resizebox{0.8\textwidth}{!}{%
\begin{tabular}{lcc}
\toprule
Failure class & Observable trigger & Policy target \\
\midrule
Wrong anchor & many neighborhoods after weak global hit & re-anchor before deeper traversal \\
Empty branch & zero-result neighborhood calls & relax relation/type or backtrack \\
Looping & repeated global/neighborhood call & suppress duplicate action; stop or reset \\
Over-broad search & many candidates, low rank movement & tighten answer type or facet \\
Premature repair & trace hint changes a good path & delay intervention until failure confidence is high \\
Budget drift & many steps without candidate gain & stop, rerank, or restart from a new anchor \\
\bottomrule
\end{tabular}}
\end{table*}

\begin{table}[h]
\centering
\caption{\textbf{From trace diagnostics to policy-learning targets.} Each row maps an observable state signal to a decision that a future controller should learn. The table is not a new result; it distills the empirical failure analysis into implementable policy targets for personalized reasoned retrieval.}
\label{tab:policy_targets}
\resizebox{0.8\columnwidth}{!}{%
\begin{tabular}{lll}
\toprule
State signal & Risk exposed by this paper & Decision to learn \\
\midrule
Profile conflicts with graph type & wrong anchor & downweight profile; search by type \\
Repeated call appears & loop or stagnation & suppress duplicate; reset or answer \\
Zero-result branch appears & empty traversal & relax relation or backtrack \\
Candidate set grows without rank gain & over-broad evidence & tighten facets or rerank \\
Budget nearly exhausted & late noisy repair & stop or return best supported list \\
\bottomrule
\end{tabular}}
\end{table}

\begin{table}[h]
\centering
\caption{\textbf{Implementation invariants.} The left column lists components that remain unchanged across all controller variants. The right column lists the only components modified by \tb. This separation is important because the work studies controller behavior, not a new graph index or new metric.}
\label{tab:invariants}
\resizebox{0.57\columnwidth}{!}{%
\begin{tabular}{lll}
\toprule
Component & Fixed across variants & Added by \tb \\
\midrule
Graph data & Same \stark SKB and node IDs & none \\
Gold labels & Same answer sets & none \\
Tools & Global search; neighborhood explore & no new tool \\
Evaluator & \hit{1}, \hit{5}, \recall{20}, \mrr & no metric change \\
Controller context & Original query and observations & profile block; trace hint \\
Logs & final answer list & trajectory counters \\
\bottomrule
\end{tabular}}
\end{table}

\begin{table}[h]
\centering
\caption{\textbf{Profile and trace fields exposed to the controller.} These fields are derived from the query and trajectory, not from gold labels. ``Purpose'' states the intended control role. The results show that exposing these fields is not sufficient; the controller must also learn when to trust or ignore them.}
\label{tab:schema}
\resizebox{0.66\columnwidth}{!}{%
\begin{tabular}{lll}
\toprule
Field group & Example fields & Intended control role \\
\midrule
Answer profile & entity type; query facets & decide broad search anchors \\
Relation hints & relation words; type constraints & choose neighborhood expansion \\
Trace counters & zero calls; repeats; budget & detect loops and empty branches \\
Candidate summary & candidate count; recent movement & stop, rerank, or restart \\
\bottomrule
\end{tabular}}
\end{table}

\begin{table}[t]
\centering
\caption{\textbf{Qualitative reading of the held-out effects.} This table translates the paired numeric deltas into the retrieval behavior they indicate. It is not a separate metric, it is more of a guide to interpreting Tables~\ref{tab:heldout} and~\ref{tab:paired}.}
\label{tab:effect_reading}
\resizebox{0.8\columnwidth}{!}{%
\begin{tabular}{lll}
\toprule
Graph & Observed pattern & Interpretation \\
\midrule
PRIME & moderate \mrr loss; high zero calls & biomedical anchoring often enters empty or wrong neighborhoods \\
MAG & largest \mrr and recall loss & wrong paper/entity anchors compound through dense relations \\
Amazon & smaller \mrr loss; recall loss remains & top plausible products survive, broader candidate coverage degrades \\
\bottomrule
\end{tabular}}
\end{table}

\begin{table}[t]
\centering
\caption{\textbf{Constrained-budget \tb probe.} The probe uses a lower step budget and stricter call caps. ``Steps'' is the average number of controller iterations actually used. The probe reduces interaction length but does not recover the unconditioned ranking quality, so the failure is not explained only by excessive trajectory length.}
\label{tab:budget}
\resizebox{0.52\columnwidth}{!}{%
\begin{tabular}{lrrrrrr}
\toprule
Graph & $n$ & \hit{1} & \hit{5} & \recall{20} & \mrr & Steps \\
\midrule
PRIME & 150 & .1200 & .2067 & .1857 & .1556 & 5.493 \\
MAG & 150 & .2800 & .4867 & .4020 & .3727 & 4.860 \\
Amazon & 150 & .4133 & .5000 & .2879 & .4511 & 4.927 \\
\bottomrule
\end{tabular}}
\end{table}

\begin{table}[t]
\centering
\caption{\textbf{Recommended reporting checklist for adaptive retrieval.} Standard rank metrics remain necessary, but they should be paired with process metrics that reveal why an agent succeeded or failed. This table summarizes the minimal diagnostics that made the \tb result interpretable.}
\label{tab:reporting}
\resizebox{0.7\columnwidth}{!}{%
\begin{tabular}{lll}
\toprule
Reported quantity & What it measures & Failure exposed \\
\midrule
Paired \mrr delta & rank movement on same IDs & hidden aggregate artifacts \\
Zero-result calls & empty branch frequency & bad anchoring or relation mismatch \\
Repeated calls & loop tendency & stalled controller state \\
Search/explore counts & breadth-depth allocation & budget misallocation \\
Constrained-budget score & robustness to shorter traces & length vs. action-quality failure \\
\bottomrule
\end{tabular}}
\end{table}

\section{Graph-Level Validation Details}
Table~\ref{tab:app_validation_graph} reports selected graph-level validation cells. The table is included to make the averages in the main text auditable. ``Global'' and ``Zero'' are trace counters and are zero for conditions without trace instrumentation.
\begin{table*}[h]
\centering
\caption{Graph-level validation details for representative controller variants.}
\label{tab:app_validation_graph}
\resizebox{0.7\columnwidth}{!}{%
\begin{tabular}{lllrrrrrrr}
\toprule
Graph & Controller & Steps & $n$ & \hit{1} & \hit{5} & \recall{20} & \mrr & Global & Zero \\
\midrule
 & Uncond. & 12 & 100 & .2100 & .3500 & .3489 & .2736 & .000 & .000 \\
PRIME & Profile & 16 & 100 & .1300 & .3600 & .3545 & .2350 & .000 & .000 \\
 & Trace & 16 & 100 & .1500 & .2600 & .2262 & .1957 & 2.260 & 1.620 \\
 & \tb & 16 & 100 & .1300 & .2000 & .1945 & .1619 & 2.130 & 1.860 \\
\midrule
 & Uncond. & 16 & 100 & .3900 & .5700 & .5127 & .4683 & .000 & .000 \\
MAG & Trace & 16 & 100 & .2900 & .4800 & .4196 & .3611 & 1.770 & 1.240 \\
 & \tb & 16 & 100 & .2500 & .4400 & .3835 & .3234 & 1.870 & 1.350 \\
\midrule
 & Uncond. & 16 & 100 & .4200 & .6300 & .4338 & .5165 & .000 & .000 \\
Amazon & Profile & 16 & 100 & .4000 & .6100 & .4149 & .4930 & .000 & .000 \\
 & \tb & 16 & 100 & .3900 & .5300 & .3214 & .4456 & 1.710 & 2.230 \\
\bottomrule
\end{tabular}}
\end{table*}

\section{Held-Out Delta Tables}
Table~\ref{tab:app_deltas} gives the held-out metric deltas used in Figure~\ref{fig:heldout_delta}. Each value is \tb minus the unconditioned agent; negative values indicate lower retrieval quality after profile/trace conditioning.

\begin{table*}[h]
\centering
\caption{Held-out aggregate deltas.}
\label{tab:app_deltas}
\resizebox{0.5\columnwidth}{!}{%
\begin{tabular}{lrrrr}
\toprule
Graph & $\Delta\hit{1}$ & $\Delta\hit{5}$ & $\Delta\recall{20}$ & $\Delta\mrr$ \\
\midrule
PRIME & -.0566 & -.0607 & -.0384 & -.0619 \\
MAG & -.2991 & -.2428 & -.2477 & -.2737 \\
Amazon & -.0406 & -.0561 & -.0950 & -.0438 \\
\bottomrule
\end{tabular}}
\end{table*}

\section{Win, Tie, and Loss Counts}
Table~\ref{tab:app_wtl} complements the paired deltas by counting whether \tb wins, ties, or loses against the unconditioned agent on each shared query.

\begin{table*}[h]
\centering
\caption{Paired win/tie/loss counts for \tb versus the unconditioned agent. Counts are computed on the same shared logged query IDs as Table~\ref{tab:paired}.}
\label{tab:app_wtl}
\resizebox{0.5\columnwidth}{!}{%
\begin{tabular}{lrrrrrr}
\toprule
Graph & MRR win & MRR tie & MRR loss & H1 win & H1 tie & H1 loss \\
\midrule
PRIME & 38 & 192 & 61 & 21 & 232 & 38 \\
MAG & 29 & 138 & 133 & 21 & 169 & 110 \\
Amazon & 43 & 190 & 64 & 25 & 235 & 37 \\
\bottomrule
\end{tabular}}
\end{table*}

\section{Trace Schema}
Each trajectory log contains the query identifier, graph name, controller variant, final answer list, tool-call sequence, observation summaries, and aggregate counters. The counters used in the main paper are computed from the raw sequence after the run; they are not hand-labeled. This schema is sufficient to reproduce per-query paired deltas, conditional repeated-call analysis, and budget-consumption summaries.

\section{Prompt Skeletons}
The exact prompts are implementation artifacts, but the following skeleton captures the logical fields used by \tb. The profile block is created before retrieval. The trace block is refreshed after each action and is intentionally limited to observable trajectory facts.

{\footnotesize
\begin{verbatim}
[QUERY]
{natural_language_question}

[QUERY PROFILE]
- predicted answer type: {entity_type}
- key facets: {facet_1, facet_2, ...}
- relation hints: {relation_words}
- retrieval risks: {ambiguity_or_missing_filter}

[TRACE SUMMARY]
- global searches used: {g_t}
- neighborhood calls used: {n_t}
- zero-result calls: {z_t}
- repeated calls: {r_t}
- remaining budget: {b_t}
- current candidate summary: {c_t}

[CONTROL INSTRUCTION]
Choose exactly one: Search, Explore, Answer, or Backtrack.
Do not repeat an earlier call unless it is necessary.
Prefer answering when additional calls are unlikely to improve rank.
\end{verbatim}}

\section{Metric Definitions}
For completeness, this section states the metric interpretation used throughout the paper. \hit{1} is one when a correct answer appears at rank one. \hit{5} is one when a correct answer appears in the top five. \recall{20} is the fraction of gold answers retrieved in the first twenty positions. \mrr is the reciprocal rank of the first correct answer. For paired deltas, each metric is computed for \tb and for the unconditioned agent on the same query ID, and then subtracted as \tb minus unconditioned.

\end{document}